\documentclass{article}
\usepackage{spconf,amsmath,epsfig}
\usepackage{multirow}
\usepackage{url}
\usepackage{booktabs}

\let\OLDthebibliography\thebibliography
\renewcommand\thebibliography[1]{
  \OLDthebibliography{#1}
  \setlength{\parskip}{0pt}
  \setlength{\itemsep}{0pt plus 0.3ex}
}

\begin{document}\sloppy

\def\x{{\mathbf x}}
\def\L{{\cal L}}

\title{An Order-Complexity Model for Aesthetic Quality Assessment of Symbolic Homophony Music Scores}
%
\name{Xin Jin, Wu Zhou, Jinyu Wang, Duo Xu, Yiqing Rong, Shuai Cui}
\address{}

\maketitle

\begin{abstract}
    Computational aesthetics evaluation has made great achievements in the field of visual arts, but the research work on music still needs to be explored. Although the existing work of music generation is very substantial, the quality of music score generated by AI is relatively poor compared with that created by human composers. The music scores created by AI are usually monotonous and devoid of emotion. Based on Birkhoff’s aesthetic measure, this paper proposes an objective quantitative evaluation method for homophony music score aesthetic quality assessment. The main contributions of our work are as follows: first, we put forward a homophony music score aesthetic model to objectively evaluate the quality of music score as a baseline model; second, we put forward eight basic music features and four music aesthetic features.
\end{abstract}
\section{Introduction}
    Computational aesthetics evaluation \cite{galanter2012computational} enables computers to make qualitative or quantitative aesthetic judgments on works of art. These works of art usually include painting, music and design. It is meaningful for computers to realize beauty because this can guide AI generatation tasks.

    Although the existing work of music generation is very mature, the quality of music score generated by AI is relatively poor compared with that created by human composers. This is probably because the essence of AI generation task is to predict the probability of the next music unit being played and the lack of prior music theory knowledge leads to the music generated by AI sounds unpleasant.

    There are three main steps in the production of pop music: composition of music score, arrangement, and finally played by the performer. We hope that the quality of score can be evaluated from the stage of music score composition, so as to eliminate the interference of different performers' performance levels on the evaluation of music score quality.
    
    Due to a lack of labeled aesthetic score on music scores like AVA \cite{murray2012ava} in the field of image aesthetic, we adopt the traditional aesthetic measure method to study the aesthetic model. Traditional aesthetic measure can analyze the beauty of objects from the perspective of information in objects.

    \begin{figure}[htbp]
        \centering
        \includegraphics[width=0.48\textwidth]{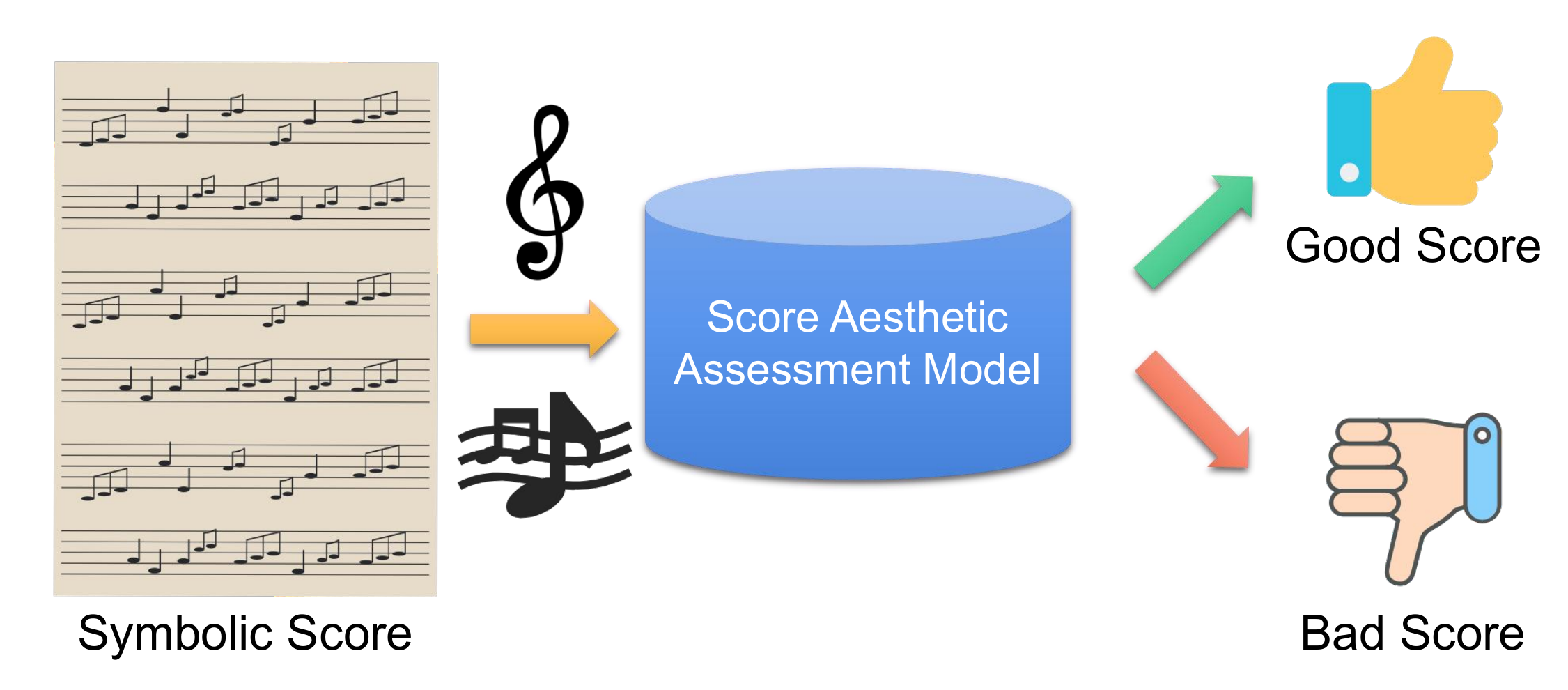}
        \caption{The quality of symbolic score can be easily evaluated through the Score Aesthetic Assessment Model (SAAM).}
        \label{fig:good_bad}
        \label{fig:good_bad}
    \end{figure}

    Our goal is to create a music score aesthetic assessment model that can objectively distinguish the good from the bad.

    In this paper, Birkhoff's method \cite{birkhoff2013aesthetic} was selected to conduct a study of aesthetic quality assessment of music score from the perspective of information theory. Birkhoff formalizes the aesthetic measure of an object into the quotient between order and complexity:

    \begin{equation}
        M=\frac{O}{C} \label{1}
    \end{equation}

   Fig 1 briefly describes the content of our work. The main contributions of our work are as follows: 
        \begin{itemize}
         \item  We put forward a score aesthetic assessment model to objectively evaluate the quality of homophony music score as a baseline score aesthetic assessment model.
        \item  We put forward and update eight basic music features and four music aesthetic features in combination with information theory and music theory.
       \end{itemize}
\begin{figure*}[htbp]
    \centering
    \includegraphics[width=\textwidth]{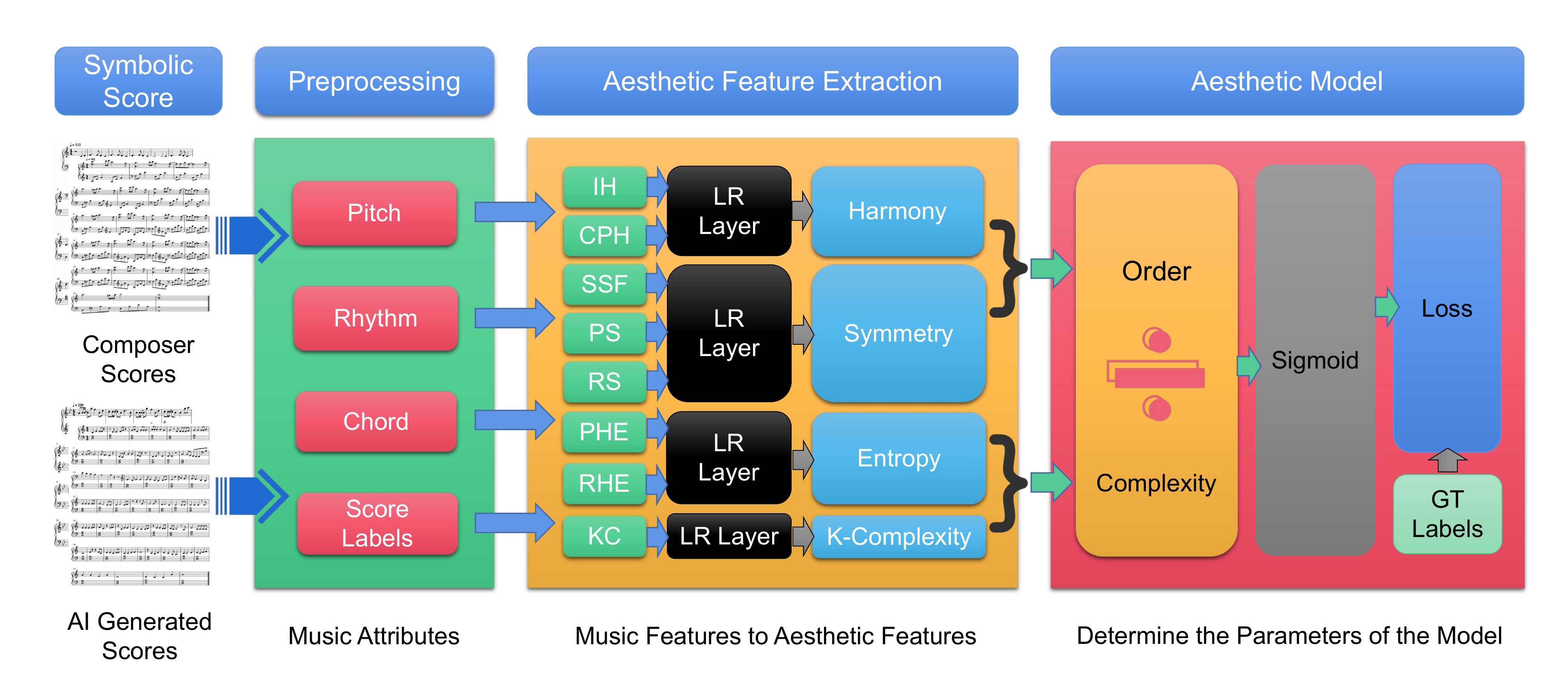}
    \caption{First, we do not tokenize the symbolic score. We extract the attributes in the symbolic score, which requires preprocessing. After preprocessing, we get the pitch, rhythm, chord attributes of the score and the label of the score as the ground truss for classification. Then, we process the music attributes and extract 8 music features (small green boxes). Next, we train the samples through four logical regression models (LR stands for logical regression) and combined to extract four aesthetic features (light blue boxes). Finally, we input the four aesthetic attributes into our model, use sigmoid function to establish the loss of its error with the ground truth, and calculate the parameters of the aesthetic model.}
    \label{fig:networknew}
    \label{fig:networknew}
\end{figure*}

\section{Related Work}
    There is only one work of music aesthetic measure using information theory, which is the aesthetic measure of audio. Audio Oracle \cite{dubnov2011audio} (AO) uses Information Rate (IR) as an aesthetic measure. However, it cannot clarify what kind of specific aesthetic intention the system has, because repetition or redundancy (proposed by IR) has essentially different meanings, interpretations and values in art, and it is questionable to use information rate as the aesthetic measure.

    There are three levels of music generation: score generation, performance generation and audio generation. The work we discuss in this section is related to music score generation. In order to make full use of music theory and study complexity, we will not discuss monophonic music, but we will discuss homophony. Homophony is a kind of multipart music, which has a melodic part and an accompanying part. In the generation of homophony music field, there are tasks such as directly generating melody and chord \cite{lim2020style}, generating melody through chord \cite{inprovrnn}, and generating chord according to melody \cite{zhu2020pop}. Although the existing generation technology is rather mature, the quality of music score created by AI is still low, which sounds monotonous and lacks emotion.

    In order to evaluate the quality of music, in the field of music generation, the evaluation part is often divided into objective evaluation and subjective evaluation. The objective evaluation is often to set some statistical metrics for the music generated by AI, and the result of objective evaluation is completely calculated by computer. Many music toolkits have objective evaluation metrics packaged for direct use, such as MV2H \cite{mcleod2018evaluating} and MusPy \cite{dong2020muspy}, etc. MV2H evaluated how many errors there are between the generated music and the ground truth. In Muspy, basic statistical metrics of symbolic score are provided. The subjective evaluation generally includes listening test and visual analysis, almost all AI generation tasks involve subjective evaluation experiments such as scoring and Turing test, which are considered necessary and essential.

\section{Score Aesthetic Assessment Model}

\subsection{Formalization of the Model}
    
    Based on Birkhoff's theory, information theory and music theory, we propose four aesthetic features: harmony, symmetry, entropy and K-Complexity. We linearly combine the order measures of molecules and the complexity measures of denominators. Detailed measures explaination will be described in Sections 3.2 and 3.3. Fig 2 shows the process of our work. The music aesthetic measure formula is as follows:

    \begin{equation}
        Aesthetic\ Measure=\frac{\omega_1H + \omega_2S + \theta_1}{\omega_3E + \omega_4K + \theta_2} \label{6}
    \end{equation}

    Where $H$ is harmony, $S$ is symmetry, $E$ is entropy and $k$ is K complexity. $\omega$ is the weight and $\theta$ is the constant.
\subsection{Order Measures}
    When objects have some characteristics of harmony, symmetry or order, they often have a certain sense of beauty. We quantify the order of music in two dimensions, harmony and symmetry. Harmony mainly calculates based on music theory knowledge, while symmetry mainly relies on some statistical information in music. Next, we use the linear combination of harmony and symmetry as the measure of order.
    
\subsubsection{Interval Harmony}

    In music, the distance between two notes is called interval. In particular, in music, when an interval is 12 semitone, we call it an octave. Interval classification can be found in the supplementary material, interval are divided into five categories.

    Mathematical and physical research shows that when two sound frequencies are a simple integer ratio, it is more pleasant to listen together. Therefore, we propose a calculation method of interval harmony, and the formula is as follows.
    
    \begin{equation}
        Interval\ Harmony=\sum_{i=1}^{12} \alpha_i * pir_i + \theta_{ih} \label{1}
    \end{equation}

    Among them, $\alpha_i$ is the weight of interval, $pir_i$ is the ratio of interval to total interval, $\theta_{ih}$ is the constant.

\subsubsection{Chord Progression Harmony}

    According to Schoenberg's theory of harmony \cite{schoenberg1983theory}, the internal chord is divided into three functional harmonies: tonic triad (T), subdominant triad (S) and dominant triad (D). An example of chord functions and series are shown in supplementary material.

    A complete harmony progression starts from the tonic triad, proceeds to the subordinate triad, proceeds to the dominant triad, and finally returns to the tonic triad to complete a complete cycle, which is called complete progression. In the usual sense, harmony progression is the connection of chords within a certain harmonic range in tonal music.
    
    There are many ways to quantify harmony progression. In this paper, we refer to the method of María \cite{navarro2020computational}. In our work, we took the average value of the progression tension to obtain a quantitative chord progress harmony. It is calculated by referring to the following formula:

    \begin{equation}
        \begin{split}
        Chord\ Progression\ Harmony = \lambda_1 d_1(T_i,T_{i-1})+ \\
        \lambda_2 d_2(T_i, T_{key}) + \lambda_3 d_3(T_i-T_{key}, T_f) + \lambda_4 c(T_i) + \\
        \lambda_5 m(T_i,P) + \lambda_6 h(T_i,P)
        \end{split}
    \end{equation}

    Where $T_i$ is the i-th chord of progression $P$, $\lambda$ is the weight. For more information of parameters $c$, $m$, $h$, see \cite{navarro2020computational}.

\subsubsection{Self Similarity Fitness}

    In the field of music generation, structure is often discussed as an important feature. Almost all music contains repetitive pieces. We will discuss the influence of repetitive structure on music aesthetics. We measure it with self similarity fitness.

    Inspired by the aesthetics of images and art \cite{al2017symmetry}, the aesthetics beauty of music comes from the symmetry in musical compositions. Therefore, in our study, we refer to Müller's fitness method \cite{MuellerJG13_StructureAnaylsis_IEEE-TASLP} to measure the degree of repetition in a piece of music. The fitness formula is shown as follow:

    \begin{equation}
        Self\ Similarity\ Fitness = 2 \cdot \frac{\bar{\sigma}(\alpha) \cdot \bar{\gamma}(\alpha)}{\bar{\sigma}(\alpha)+\bar{\gamma}(\alpha)}
    \end{equation}

    Both $\bar{\sigma}(\alpha)$ and $\bar{\gamma}(\alpha)$ are related to a concept defined by Müller's method \cite{MuellerJG13_StructureAnaylsis_IEEE-TASLP}.
\subsubsection{Skewness}

    Skewness is a concept proposed by jSymbolic \cite{mckay2018jsymbolic}. It proposes that both pitch and rhythm of music have the concept of skewness. The notes in music cannot lack pitch and rhythm. Skewness describes how asymmetrical the pitch / rhythm is to either the left or the right of the mean pitch / rhythm value.

    The features are extracted based on jSymbolic, the calculation formula of pitch and rhythm skewness are not specifically described here. We combine pitch skewness and rhythm skewness linearly to get the formula of skewness:

    \begin{equation}
        Skewness = \beta_1 * PS + \beta_2 * RS + \theta_{sk}
    \end{equation}

    Where $PS$ is pitch skewness, $RS$ is rhythm skewness, $\beta$ represents their weight and $\theta_{sk}$ is a constant.
\subsection{Complexity Measures}

    Bense \cite{bense1960programmierung} first uses Birkhoff's aesthetic measure formula to calculate aesthetics. They adapt statistical measure of information in aesthetic objects and believe that the objective measure of aesthetic objects is related to the complexity of objects. Their idea has to use information theory, and entropy is the core of it. Our aesthetic measure method considers two features: Shannon entropy and Kolmogorov complexity.
    
\subsubsection{Shannon Entropy}

    Let $\Omega$ be a finite set, and $X$ be a random variable. The value $x$ in $\Omega$ has a distribution $p(x)=Pr[X=x]$. The Shannon entropy $H(X)$ of random variable $X$ is defined as follows:

    \begin{equation}
        H(X)=-\sum_{x\in\Omega}p(x)\log(x)
    \end{equation}

    The Shannon entropy $H(X)$ measures the average uncertainty of random variable $X$, which is widely used to evaluate the degree of chaos in the internal state of a system. In order to calculate the entropy of music, it is necessary to obtain the music attribute histogram.

    Pitch and rhythm are the two basic elements of music. In our method, we consider pitch entropy and rhythm histogram entropy, and take their linear combination as the measure of entropy. This is used to describe the uncertainty in music. Our entropy formula is as follows:

    \begin{equation}
        Entropy=\eta_1 * PHE + \eta_2 * RHE + \theta_e
    \end{equation}

    Where $PHE$ and $RHS$ are pitch and rhythm histogram entropy, $\eta$ represents their weight and $\theta_e$ is a constant.
\subsubsection{Kolmogorov Complexity}

    For a string $s$, Kolmogorov complexity $K(s)$ of the string $s$ refers to the shortest program to calculate the string $s$ on a computer. In essence, the Kolmogorov complexity of a string is the length of the final compressed version of the string.  Then, we use the linear combination of entropy and Kolmogorov complexity as a measure of complexity.
    
    Aesthetically speaking, redundancy makes people feel dull, resulting in negative emotions. According to the definition of Kolmogorov complexity, we also refer to the method of using Kolmogorov complexity in image aesthetics \cite{rigau2008informational}.
    
    We believe that Kolmogorov complexity in music is also computable. It is actually the lossless compression ratio of music, which can be formalized as the following formula:

    \begin{align}
        Kolmogorov\ Complexity=\frac{NH_{m} - K}{NH_{m}} \label{1}
    \end{align}

    Where $NH_{m}$ is information content of a music, and $K$ is the simplest music information after compression.
    \begin{figure*}[htbp]
        \centering
        \includegraphics[width=1\textwidth]{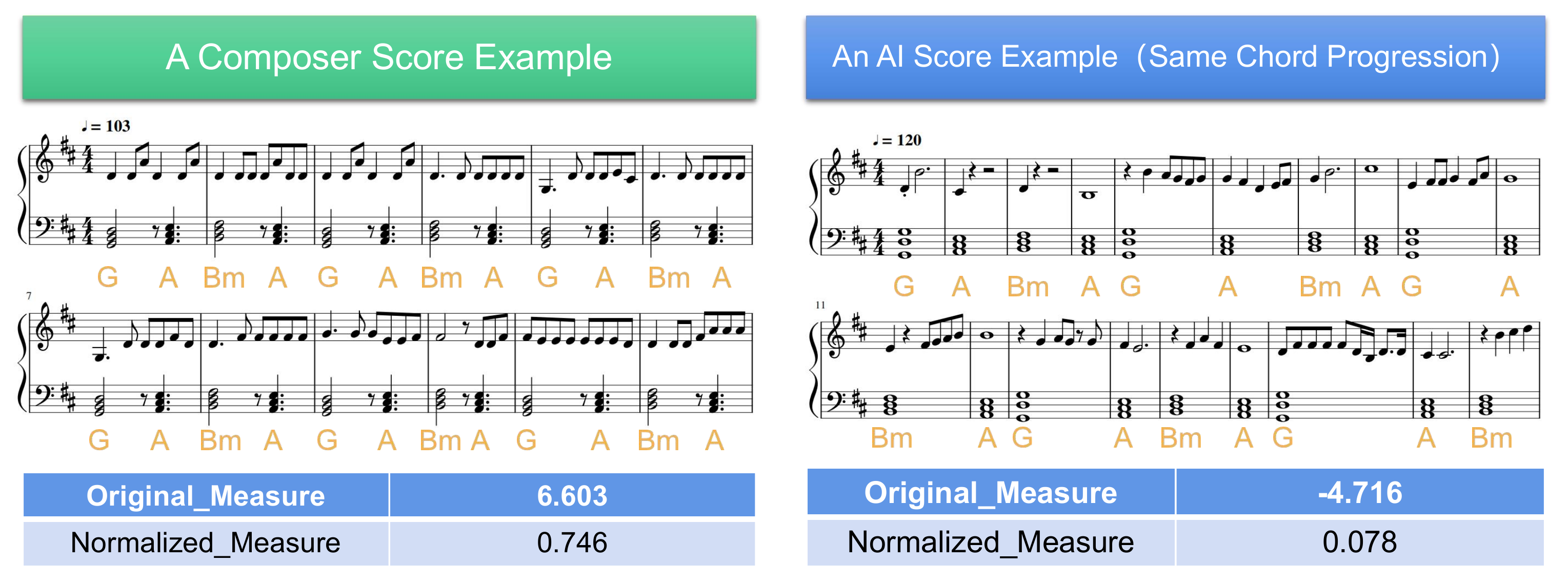}
        \caption{As can be seen from the figure, the melody of composer score is in order, and it matches the chord progression very well. The melody of AI score seems very random and low-quality, which is specifically reflected in the irregular appearance of the rest and the melody do not follow the tension of the chord progression. \textbf{Note: The full score of normalized measure is 1.}}
    \end{figure*}
    
\section{Implementation}

\subsection{Datasets}
    There are many datasets of pop music, such as POP909 \cite{wang2020pop909} and Wikifonia\footnotemark{}. But we don't use them due to some reasons.
    
    All music data formats in POP909 dataset are performance midi, which will cause different performers' aesthetic impact on the same music score. Although Wikifonia is a score dataset (single track), it can't be used by order measure.

    \footnotetext{http://www.wikifonia.org/}
    
    In order to eliminate the performance differences of different performers, objectively measure the aesthetic value of a piece of music, we finally download the scores of 100 pop songs on the Musescore\footnotemark{} website. Then we extract the chords of music scores in Musescore for generating 100 music scores. In datail, we take chord progression from the scores of Musescore, and determine scores' key signature (tonic) as the input parameter, then give them to Magenta's inprov\_rnn \cite{inprovrnn} enables it to generate music according to chord progression and key signature, which has the advantage of controlling the time length of two pairs of datasets to be approximately the same. The scores in Musescore are created by composers as positive samples, while the scores created by Magenta are generated by AI as negative samples.

    \footnotetext{https://musescore.org/}

    Since we use cross validation, we do not set validation sets. We split the dataset at a ratio of 7:3 for training testing.
\subsection{Preprocessing \& Computing Aesthetic Features}
    We use music21 \cite{cuthbert2010music21} toolkit to load music scores. Music21 has note and chord attributes, which can easily obtain the information of music score and make calculations.

    Firstly, we obtain the pitch histogram and rhythm histogram in the music score to calculate entropy. Then, we get all the intervals by calculating the note events that occur at the same time, and obtain a histogram. So, the histogram entropy of pitch and rhythm and interval harmony can be calculated.

    Secondly, we try to get all the chords and the key signature in the score to calculate the chord progress harmony. The chord progress and the key signature will be saved in the json format and subsequently input to the pretrained model.
    
    Thirdly, we use Musescore3 to batch process composer scores and AI scores into score midi. This is to facilitate the use of jSymbolic \cite{mckay2018jsymbolic} to extract features. The conversion of xml to score midi will not lose the score information. In this way, we get pitch and rhythm skewness.

    Fourthly, we use Musescore3 to render the music score into audio, which controls that all music scores are played by Musescore3 to ensure that the music score information is lossless. Then we use the wav file format to calculate the self similarity fitness. Then, we refer to Monkey's Audio's\footnotemark{}\footnotetext{https://www.monkeysaudio.com/} lossless compression method to compress music into ape format, so as to calculate Kolmogorov complexity.

    Fifthly, calculating the values of 8 basic music features (refer to Table 1), we will normalize them. Next, we use the method of logical regression to confirm the values of Harmony, Symmetry, Entropy, and Kolmogorov complexity.

    Finally, we take the four normalized aesthetic features as inputs, use sigmoid function to map the aesthetic measure to 0 and 1 for classification, and use cross-entropy loss to make loss function, confirming the parameters of the aesthetic model in the way of gradient decline. We set the learning rate to 0.01. After 1000 iterations, the loss function converges.

\section{Experiments}

\subsection{Turing Test}

    We need to verify whether people really think the music created by composers is more beautiful than the music generated by AI. So we did the Turing test to see if people can really distinguish between composer music and AI music.

    We randomly sampled 10 pieces from the music created by the composer and the music generated by AI respectively, with each piece lasting about 15 seconds. Volunteers participating in the Turing test need to identify which music is created by the composer and which music is generated by AI in these 10 pairs of pieces. The volunteer also needs to choose the one in each pair that he thinks is more beautiful.

    A total of 15 volunteers participate in the Turing test. Among 300 samples, their classification accuracy of music is 91.3\% (274). Assuming that the music created by the composer is more aesthetic, 87\% (261) of the samples are correctly classified. This proves that our assumption is correct.

\subsection{Results \& Discussion}

    Since the music created by the composer has a higher aesthetic feeling than the music generated by AI, we let the machine learn the aesthetic score according to the label value. This is essentially a binary classification problem.

    \begin{figure}[!htbp]
        \centering
        \includegraphics[width=0.352\textwidth]{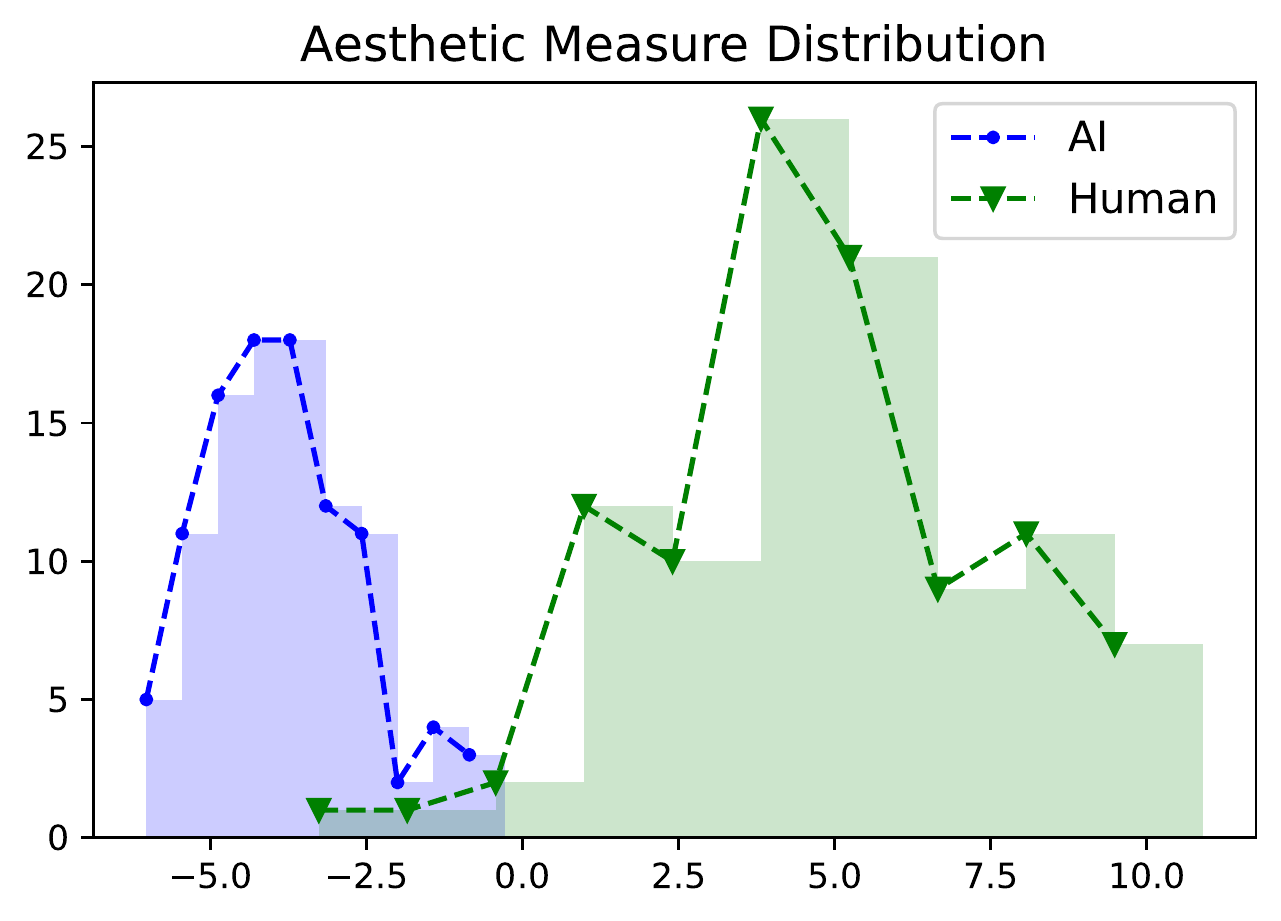}
        \caption{The intersection of the distributions is quite small, showing that our model can distinguish between AI scores and composer scores very well.}
        \label{fig:distribution}
        \label{fig:distribution}
    \end{figure}
    
    After the weight is obtained by gradient descent, we bring the weight into the aesthetic model to view the distribution of the aesthetic model. The distribution is shown in Fig 4.

    We use precision and F1-Measure as our metric to test our model. The precision of our model on the test set is 93.3\%, and the F1-Measure is 90.9\%. This proves that our model is valid. Fig 3 shows an example of score comparison.

    \begin{table}[htbp]
    \centering
        \resizebox{1\columnwidth}{!}{
        \begin{tabular}{c|c|c|c|c|c|c|c|c}
        \hline
            \textbf{Dataset} & \textbf{IH} & \textbf{CPH} & \textbf{SSF} & \textbf{PS} & \textbf{RS} & \textbf{PHE} & \textbf{RHE} & \textbf{KC} \\ \hline
            AI & 0.97 & 1.98 & 0.06 & \textbf{0.49} & 2.16 & \textbf{1.34} & \textbf{1.60} & \textbf{0.69} \\ \hline
            Composer & \textbf{1.40} & \textbf{2.04} & \textbf{0.19} & 0.53 & \textbf{0.99} & 2.16 & 1.79 & 0.73 \\ \hline
        \end{tabular}
    }
    \caption{The similar CPH value shows the rationality of the AI dataset generated based on chords. \textbf{More detailed distribution comparison can be found in the supplementary material.} The results in bold have higher aesthetic scores.}
\end{table}
    
    Table 1 shows the calculation results of 8 features without normalization. We can make the following analysis from it. Whether IH or CPH, the scores of composer is obviously better than that of AI. This is also in line with the music theory. Considering Symmetry, the SSF of composer is obviously higher than that of AI. This is because the music created by AI is too random and often has no repetitive pieces, which proves beauty is related to repetition. Although the PS of AI is smaller than that of the composer, the difference is not much. The RS of the composer is obviously smaller than that of AI. This is because composers tend to create more regular rhythm and pitch than AI. When it comes to entropy, the values of PE and HE are obviously higher in composer score than in AI score. Although this is contrary to Birkhoff's aesthetic measure, it is also reasonable. This is because composer tends to add some changes to pitch and rhythm, while music generated by AI does not change much around the tonic. As for K-complexity, the difference between composer's scores and AI's scores is not significant, but it can be seen from the table that the music created by AI is relatively low in compression and complexity. In conclusion, moderate order and complexity quotient can quantify aesthetic feeling to a certain extent.

\subsection{Ablation Study}

    We conduct ablation experiments to remove harmony, symmetry, entropy and K-complexity respectively to train four different models. We compare them with the original model.
    
    As shown in Fig 5, we observe the ROC curves of the original model and four models with one aesthetic feature removed respectively, and obtain their AUC values. The AUC value of our model is 0.93, which is obviously higher than that of the four models without aesthetic features.

    If harmony is removed, the AUC value is only 0.77, which shows the importance of harmony and further proves that music theory plays a very important role in music aesthetics. If symmetry is removed, the AUC value of the model is 0.85, which indicates that symmetry may contribute slightly less to aesthetics than harmony. As for entropy, it is obviously important, but K-complexity seems not.
    \begin{figure}[!htbp]
        \centering
        \includegraphics[width=0.38\textwidth]{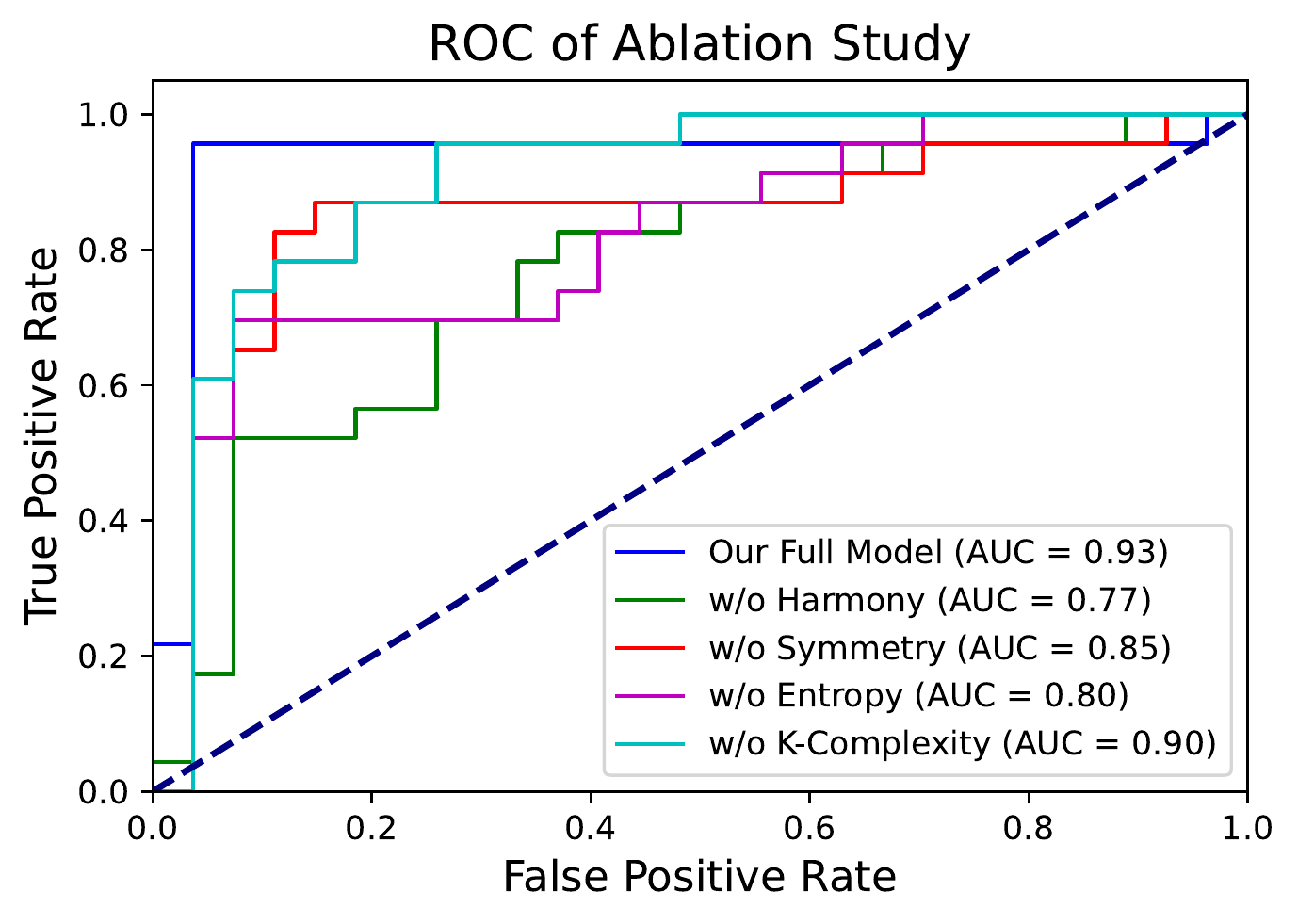}
        \caption{The ROC curves are serrated due to lack of samples.}
        \label{fig:ablation}
        \label{fig:ablation}
    \end{figure}

\section{Conclusion}
    In summary, we propose a score aesthetic assessment model using Birkhoff's aesthetic measure to quantify a score's aesthetic. We also discover four categories of music aesthetic features, totaling eight basic aesthetic features. We have made some contributions to improve the quality of music scores. This might be helpful for music score quality assessment. However, our method still has shortcomings, for instance we have not taken the relationship between creativity and musical aesthetics into consideration. The aesthetic study of music audio quality assessment is worth exploring in the future.

\bibliographystyle{IEEEbib}
\bibliography{icme2023template}
\end{document}